\documentclass[prb,twocolumn,amsmath,amsfonts,floatfix,showpacs,letterpaper]{revtex4}

\pdfoutput=1

\usepackage{graphicx}
\usepackage{ifpdf}

\DeclareMathOperator{\Tr}{Tr}

\newcommand{\beq}[1]{\begin{equation}\label{#1}}
\newcommand{\eeq}{\end{equation}}
\newcommand{\refeq}[1]{Eq.~(\ref{#1})}

\newcommand{\beqm}[1]{\begin{multline}\label{#1}}
\newcommand{\eeqm}{\end{multline}}

\newcommand{\punc}[1]{\,{\text{#1}}}
\newcommand{\sub}[1]{_{\mathrm{#1}}}

\newcommand{\Z}{{\mathcal{Z}}}
\newcommand{\T}{{\mathcal{T}}}
\newcommand{\Ham}{{\mathcal{H}}}
\newcommand{\Act}{{\mathcal{S}}}
\newcommand{\Lag}{{\mathcal{L}}}
\newcommand{\D}{{\mathcal{D}}}
\newcommand{\C}{{\mathcal{C}}}

\newcommand{\stateenergy}{E}
\newcommand{\clH}{\mathcal{E}}
\newcommand{\quH}{\mathcal{H}}

\newcommand{\timeordered}{{\mathbb{T}_\tau}}

\newcommand{\phInvert}{\mathcal{I}\sub{ph}}
\newcommand{\xReflect}{\mathcal{R}_x}
\newcommand{\yReflect}{\mathcal{R}_y}
\newcommand{\yzReflect}{\mathcal{I}_y}
\newcommand{\xTrans}{\mathcal{K}_x}
\newcommand{\yTrans}{\mathcal{K}_y}
\newcommand{\zTrans}{\mathcal{K}_z}

\newcommand{\boH}{{\mathbf{H}}}
\newcommand{\boeta}{{\boldsymbol{\eta}}}
\newcommand{\boalpha}{{\boldsymbol{\alpha}}}
\newcommand{\boalt}{{\tilde{\boldsymbol{\alpha}}}}
\newcommand{\boaltbar}{{\bar{\tilde{\boldsymbol{\alpha}}}}}

\newcommand{\at}{{\tilde{a}}}
\newcommand{\atb}{{\bar{\tilde{a}}}}

\newcommand{\ns}{^{\phantom{*}}}
\newcommand{\nd}{^{\phantom{\dagger}}}

\newcommand{\Sv}{{\mathbf{S}}}
\newcommand{\Bv}{{\mathbf{B}}}
\newcommand{\Av}{{\mathbf{A}}}

\newcommand{\sublattice}[1]{\ensuremath{\mathsf{#1}}}

\newcommand{\ev}{\mathbf{e}}
\newcommand{\ea}{{\mathbf{e}_\sublattice{a}}}
\newcommand{\eb}{{\mathbf{e}_\sublattice{b}}}

\newcommand{\kv}{{\mathbf{k}}}
\newcommand{\Kv}{{\mathbf{K}}}
\newcommand{\rv}{{\mathbf{r}}}
\newcommand{\rvp}{{\mathbf{r}'}}
\newcommand{\zerov}{{\mathbf{0}}}

\newcommand{\ee}{\mathrm{e}}
\newcommand{\ii}{\mathrm{i}}
\newcommand{\dd}{\mathrm{d}}

\newcommand{\Order}[1]{{\mathcal{O}(#1)}}

\newcommand{\ket}[1]{{|#1\rangle}}
\newcommand{\bra}[1]{{\langle#1|}}

\usepackage{color}

\ifpdf

\newcommand{\putinscaledfigure}[1]{\begin{center}\includegraphics[width=\columnwidth]{#1}\end{center}}
\else

\newcommand{\putinscaledfigure}[1]{\fbox{#1}}
\fi

\begin{document}

\title{Classical--Quantum Mappings for Geometrically Frustrated Systems:\\Spin Ice in a $[100]$ Field}

\author{Stephen Powell}
\author{J.\ T.\ Chalker}
\affiliation{Theoretical Physics, Oxford University, 1 Keble Road, Oxford, OX1 3NP, United Kingdom}

\begin{abstract}
Certain classical statistical systems with strong local constraints are known to exhibit Coulomb phases, where long-range correlation functions have power-law forms. Continuous transitions from these into ordered phases cannot be described by a na\"\i ve application of the Landau-Ginzburg-Wilson theory, since neither phase is thermally disordered. We present an alternative approach to a critical theory for such systems, based on a mapping to a quantum problem in one fewer spatial dimensions. We apply this method to spin ice, a magnetic material with geometrical frustration, which exhibits a Coulomb phase and a continuous transition to an ordered state in the presence of a magnetic field applied in the $[100]$ direction.
\end{abstract}


\maketitle

\section{Introduction}

Classical statistical systems with Coulomb phases, where long-range correlation functions have power-law forms, have been of great interest in recent years. Of particular interest are continuous transitions from Coulomb to ordered phases,\cite{Alet,Jaubert,Pickles} which are {\it prima facie} incompatible with the standard Landau-Ginzburg-Wilson theory, since neither phase is simply thermally disordered.

Various classical systems exhibiting power-law correlation functions---such as geometrically frustrated anti\-ferromagnets,\cite{Anderson,Pickles} crystalline water ice\cite{Pauling,BramwellGingras} and close-packed dimers on square and cubic lattices\cite{Fowler,Alet}---have in common local constraints arising from dominant terms in the microscopic Hamiltonian. In an effective description of the Coulomb phase, the local degrees of freedom are replaced by a continuum field, with the constraints replaced by a requirement that the field be divergenceless. The power-law behaviour is then understood in terms of the correlations of this solenoidal field.\cite{Huse,Isakov1,Henley}

This coarse-grained picture does not, however, allow a description of certain ordered phases where the discrete nature of the local degrees of freedom is crucial, and therefore cannot describe transitions into these phases. These include states with saturated magnetic moments, as well as ordered phases in dimer models with hard-core repulsion between dimers.\cite{Alet}

Here we will describe an alternative approach, which leads to an effective critical theory for continuous transitions from Coulomb phases to such ordered phases. It is based on the standard mapping between a thermal phase transition in $d$ spatial dimensions and a quantum phase transition at zero temperature in $d-1$ dimensions. The local degrees of freedom are represented by hard-core bosons, allowing their discrete nature to be treated exactly, while the power-law correlations in the Coulomb phase are reproduced by the fluctuations of the Goldstone mode that appears when the bosons condense.

One clear disadvantage of the method is that it necessarily treats the system anisotropically, with one of the spatial dimensions being mapped to the quantum imaginary time, but it should be noted that space-time isotropy is often restored in the continuum limit of various quantum models.

In a forthcoming work, we will present the application of this approach to the ordering transition of close-packed dimers on the cubic lattice.\cite{Alet} Here, we address a model of spin ice\cite{BramwellGingras,Isakov} in an applied magnetic field aligned along the $[100]$ crystallographic direction.\cite{Jaubert} This system is convenient firstly because the mapping to the quantum problem takes a particularly transparent form, and secondly because the isotropy of the classical system is already broken by the magnetic field. The problem is also interesting in its own right because the phase transition is produced by an external field, and because, as we explain in Section~\ref{SecStrings}, it is an unusual example of a Kasteleyn transition\cite{Kasteleyn,Bhattacharjee} in three dimensions.

A brief account of the application of a simplified version of this method to spin ice has appeared elsewhere,\cite{Jaubert} along with results from Monte Carlo simulations. It was shown there that a $[100]$ magnetic field can cause a Kasteleyn transition from the Coulomb phase to an ordered phase where the spins are aligned with the field and fluctuations are frozen. For the corresponding quantum problem, this becomes a quantum phase transition from a Bose condensate to a vacuum state.

Here we present the method in full, treating properly the crystal structure of spin ice, and show that doing so leads to an effective quantum Hamiltonian with nonhermitian directed-hopping terms. We demonstrate that these terms are crucial to understanding the Coulomb phase correlation functions via the quantum mapping. At the transition, however, we show they have no effect, so that the standard critical theory for the vacuum transition of two-dimensional bosons applies, as conjectured in Ref.~\onlinecite{Jaubert}.

In the remainder of this section, we will present the Hamiltonian for spin ice and review how, in the absence of an applied magnetic field, the correlation functions in the Coulomb phase can be understood in terms of an effective solenoidal field. In Section~\ref{SecBosons}, we describe in detail the mapping to a quantum problem and the general form of the Hamiltonian that results, while in Section~\ref{SecContinuumTheory} we find the continuum action that corresponds to this Hamiltonian. This allows us to obtain the critical theory for the ordering transition and also to show that the quantum model correctly reproduces the correlation functions within the Coulomb phase. Finally, in Section~\ref{SecLargeS}, we show that these same correlation functions can be found directly from the microscopic quantum Hamiltonian, by treating the hard-core bosons as $S=\frac{1}{2}$ spins and taking the lowest order in a spin-wave expansion.

\subsection{Spin ice: Hamiltonian}
\label{SecSpinIce}

\begin{figure}
\putinscaledfigure{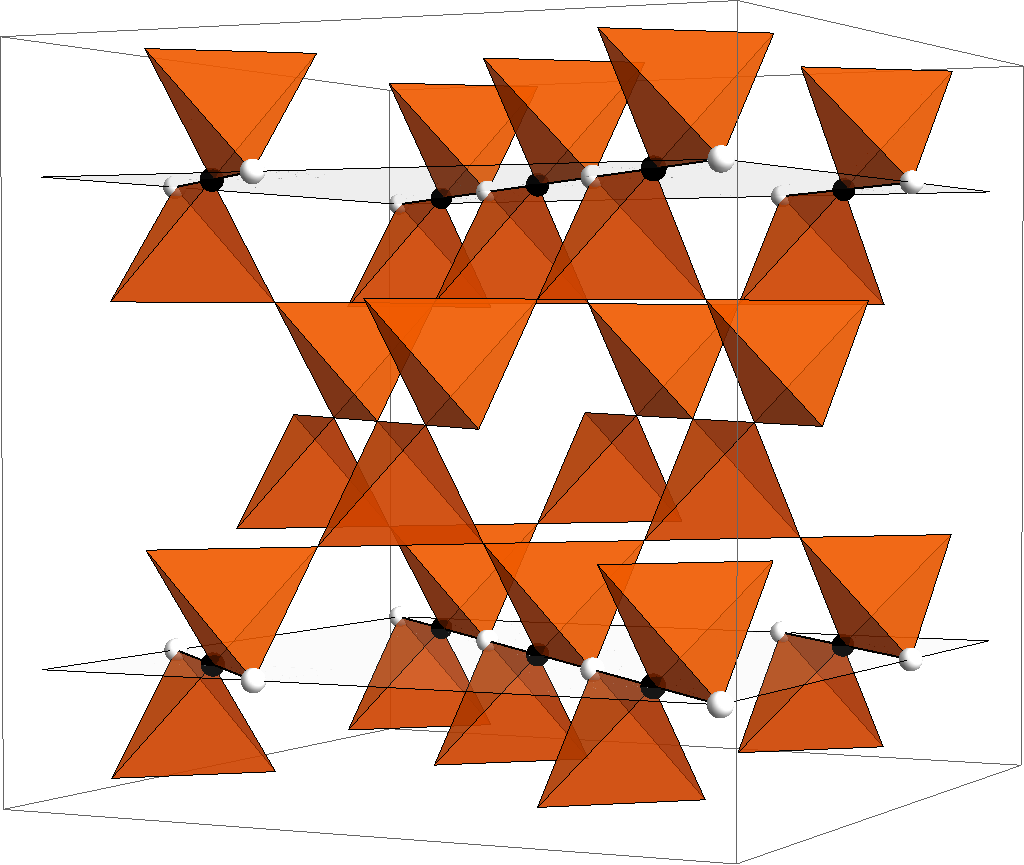}
\caption{\label{PyrochloreWith2DPlanes}A section of the pyrochlore lattice, with the $[100]$ crystallographic direction shown vertically. The sites of the lattice are at the vertices of the tetrahedra and the two $(100)$ planes shown are one possible choice to define the mapping to a quantum problem described in Section~\ref{SecBosons}. The points shown on these planes indicate the lattice sites for the quantum bosons, with black and white points respectively indicating the two distinct sublattices. We define the $[011]$ direction as parallel to the bold lines drawn diagonally across the planes, connecting nearest neighbour sites.}
\end{figure}

The spin ice compounds consist of ions with large magnetic moments ($\sim 10$ Bohr magnetons) arranged on the sites $i$ of a pyrochlore lattice (see Figure~\ref{PyrochloreWith2DPlanes}), with a strong crystal field favouring alignment of the spins along the lines joining the centres of neighbouring tetrahedra.\cite{BramwellGingras} At low temperatures, each moment $\Sv _i$ is therefore constrained to take the values $\Sv _i = S_i \ev _i$, where $S_i = \pm 1$ is an effective classical Ising spin and $\ev _i$ is a unit vector joining the centres of two neighbouring tetrahedra. (The centres of the tetrahedra form a diamond lattice, and we choose the convention where $\ev _i$ always points from a certain diamond sublattice to the other.)

In the spin ice materials $\mathrm{Ho}_2 \mathrm{Ti}_2 \mathrm{O}_7$ and $\mathrm{Dy}_2 \mathrm{Ti}_2 \mathrm{O}_7$, the largest interaction terms are long-range dipolar forces between the magnetic moments.\cite{BramwellGingras} It has been shown,\cite{Isakov} however, that the low-energy states of dipolar spin ice have a `projective equivalence' to those of a model with purely short-range ferromagnetic interactions, and we will work with this simplified model. Despite being ferromagnetic, this model is frustrated, and the ground-state degeneracy, determined experimentally by low-temperature measurements of the entropy,\cite{Ramirez} is in fact identical to that in the cubic phase of water ice.

We treat spin ice in the presence of a field $h$ along the $[100]$ crystallographic direction of the pyrochlore lattice (the vertical direction in Figure~\ref{PyrochloreWith2DPlanes}), whose effect is much smaller than the interaction between the spins. The field then acts as a perturbation within the manifold of degenerate ground states of the $h=0$ problem. The Hamiltonian can be written
\beq{SpinH}
\begin{split}
\clH &= -J \sum_{\langle i, j \rangle} \Sv_i \cdot \Sv_j - h \hat{\mathbf{n}}_{[100]} \cdot \sum_i \Sv_i\\
&= J\sub{eff} \sum_{\langle i, j \rangle} S_i S_j - h\sub{eff} \sum_i (-1)^{z_i} S_i
\punc{,}
\end{split}
\eeq
where $\langle i, j\rangle$ denotes that the sum is over nearest-neighbour sites $i$ and $j$, and $\hat{\mathbf{n}}_{[100]}$ is a unit vector in the $[100]$ direction, which we take as the $z$-axis. In the second equation, we have made use of the constraint $\Sv _i = S_i \ev _i$, and $z_i$ denotes the $z$-component of the position of site $i$. Note that, in terms of the Ising degrees of freedrom $S_i$, the effective coupling $J\sub{eff} = \frac{1}{3}J$ is antiferromagnetic and $h\sub{eff} = \frac{1}{\sqrt{3}}h$ is an effective staggered field. Throughout, we will consider low temperature $T$ and weak field, taking $h,T \ll J$.

For $h = 0$, the energy is minimized by any configuration where every tetrahedron of the pyrochlore lattice has two spins pointing inwards and two pointing outwards; such configurations are said to obey the `ice rules'.\cite{BramwellGingras} The number of such states grows exponentially with system volume, and so there is no ordering even for $T \ll J$. The long-distance spin--spin correlation functions have a power-law form, which can be found using a coarse-grained solenoidal field, as we will review in Section~\ref{SecCoulombPhase},

In the limit of large $h$, the Ising spins take the values that maximize the projection of the magnetic moment along the $[100]$ direction. This results in the `top' two spins in every tetrahedron (where $[100]$ is defined as upward) pointing outwards, while the `bottom' two point inwards. This is consistent with the ice rules and hence this state is the exact ground state for $h > 0$.

As argued in Ref.~\onlinecite{Jaubert}, there is an ordering transition for $T \sim h$ in the limit of large $J$, with the ice rules applying on both sides. One cannot straightforwardly write down a critical action of the Landau-Ginzburg-Wilson type for this strongly constrained transition.\cite{Jaubert} Instead, we use an approach that relies on a mapping from this classical statistical problem to a quantum problem, which we describe in Section~\ref{SecBosons}.

Away from the infinite $J/T$ limit, the ice rules are broken on a finite density of tetrahedra by thermal fluctuations. In this case, the phase transition is replaced by a sharp crossover, which becomes more rounded off as $J/T$ is increased (see the numerical results of Ref.~\onlinecite{Jaubert}, and in particular Fig.~2).

\subsection{Coulomb phase}
\label{SecCoulombPhase}

We first review a method for describing the behaviour on the disordered side of the transition.\cite{Isakov1} We consider a coarse-grained version of the theory, where the spins $\Sv _i$ on lattice sites $i$ are replaced by a continuum field $\Bv(\rv)$, defined for all points in space $\rv$. While this gives the appropriate behaviour deep within the disordered phase, it does not incorporate the fixed spin length and so cannot be used to describe the transition to the ordered state.

The ice rules require that the sum of $\Sv _i$ over sites in any tetrahedron should vanish, which leads in the continuum limit to the constraint that $\Bv$ should be divergenceless: $\nabla \cdot \Bv = 0$. Defining the gauge field $\Av(\rv)$ by $\Bv = \nabla \times \Av$, the only allowed continuum action respecting both the spatial symmetries and gauge invariance is
\beq{ActionForB}
\Act _\Bv \sim \int \dd^3 \rv \: (\nabla \times \Av)^2 + \cdots\punc{,}
\eeq
where the omitted terms (involving higher derivatives) are irrelevant in the renormalization group sense.

The correlation functions of the field $\Bv$ are therefore given by the standard dipolar correlation functions,
\beq{BkCorrFunc}
\langle B_i(\kv) B_j(\kv) \rangle \sim \delta_{ij} - \frac{k_i k_j}{|\kv|^2}
\eeq
in momentum space and hence
\beq{BrCorrFunc}
\langle B_i(\rv) B_j(\zerov) \rangle \sim \frac{3 r_i r_j - |\rv|^2\delta_{ij}}{|\rv|^5}
\eeq
in position space.\cite{Isakov1} The long-range correlation functions of the spins $\Sv _i$ have identical forms.

To determine the correlations of the effective Ising spins $S_i$, we must take account of the four-site basis of the pyrochlore lattice. The correlation function decreases with increasing separation $\rv$ as $|\rv|^{-3}$ and depends in sign and magnitude on the orientation of $\rv$ and the choice of sites from the basis. As we explain below, we focus on the subset of sites that lie on a particular set of (100) planes illustrated in Figure~\ref{PyrochloreWith2DPlanes}. The sites lying in such planes are shown as the points in the figure, and form two sublattices, which we label $\sublattice{a}$ (black) and $\sublattice{b}$ (white).

The directions along which the orientations of these spins are constrained are specified by the unit vectors $\ea$ and $\eb$, given by
\begin{align}
\ea &= \hat{\mathbf{n}}_{[100]} \sin\theta + \hat{\mathbf{n}}_{[011]} \cos\theta\\
\eb &= \hat{\mathbf{n}}_{[100]} \sin\theta - \hat{\mathbf{n}}_{[011]} \cos\theta \punc{,}
\end{align}
where $\tan\theta = 1/\sqrt{2}$ and $\hat{\mathbf{n}}_{[011]}$ is a unit vector in the $[011]$ direction, which is chosen parallel to the line joining neighbouring $\sublattice{a}$- and $\sublattice{b}$-sites. The spin $\Sv_{\sublattice{a}}$ at a site on sublattice $\sublattice{a}$ can be written as $\Sv_{\sublattice{a}} = S_{\sublattice{a}} \ea$ (and similarly for $\sublattice{b}$).

We therefore find that the long-range spin--spin correlation function (in momentum space) for two spins on sublattice $\sublattice{a}$ is given by
\beq{aaCorrFunc}
\langle S_{\sublattice{a}}S_{\sublattice{a}}\rangle \sim \frac{(k_{[100]} + \sqrt{2}k_{[011]})^2}{|\kv|^2}\punc{,}
\eeq
while that for two spins on opposite sublattices is
\beq{abCorrFunc}
\langle S_{\sublattice{a}}S_{\sublattice{b}}\rangle \sim \frac{k_{[100]}^2 - 2k_{[011]}^2}{|\kv|^2}\punc{.}
\eeq
(In both cases, we have omitted a constant term, which upon Fourier transforming to position space becomes a delta function at the origin, and therefore has no effect on long-distance properties.)

Note that the correlation function for sites within the $\sublattice{a}$-sublattice is maximal along a line $k_{[100]} = \sqrt{2}k_{[011]}$ in momentum space, parallel to the line (in real space) joining the centres of tetrahedra that meet at $\sublattice{a}$-sites. By contrast, the correlation function between $\sublattice{a}$- and $\sublattice{b}$-sites has no such directionality, being symmetric in $\pm k_{[011]}$.

In Section~\ref{SecCondensedPhase}, we will show that the same form for the correlation functions results from using the mapping to a quantum problem: see Eqs.~(\ref{nnCorrelation1}) to (\ref{nnCorrelation3}).

\subsection{String picture}
\label{SecStrings}

\begin{figure}
\putinscaledfigure{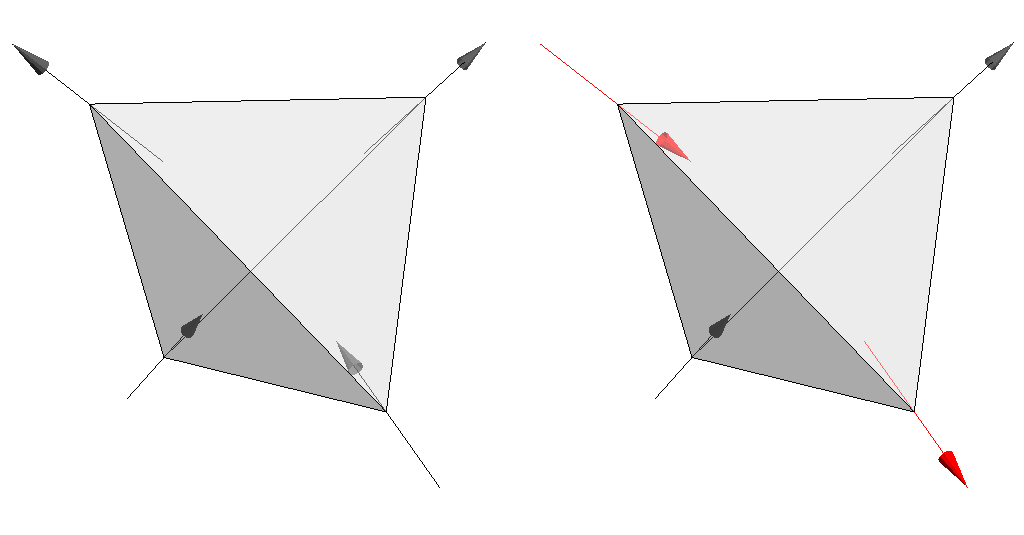}
\caption{\label{Tetrahedra}Two possible arrangements of the spins on a single tetrahedron of spin ice, where, as described in Section~\ref{SecSpinIce}, a strong crystal field constrains the spins to point either directly towards or away from the centre of each tetrahedron. Both configurations have two spins pointing outwards and two inwards and so minimize the ferromagnetic interaction energy; they are said to obey the `ice rules'. The arrangement on the left shows the lowest-energy configuration in the presence of an applied magnetic field in the $[100]$ direction (chosen vertically upwards, as in Figure~\ref{PyrochloreWith2DPlanes}), with the spins maximally aligned along the field. On the right-hand side, one of the first excited states is shown, with two spins flipped with respect to the ground state. In order to obey the ice rules, there must be the same number of flipped spins at the top of the tetrahedron as at the bottom, and so the number of flipped spins is the same in each $(100)$ plane of the pyrochlore lattice.}
\end{figure}

We now consider the ordered phase, where the coarse-grained description is no longer appropriate, and the transition from this phase into the Coulomb phase. For $T \ll h \ll J$, all moments are as aligned with the applied field as possible, given the strong crystal field. As noted above, this implies that the top two spins in every tetrahedron point out and the bottom two point in, as shown on the left-hand side of Figure~\ref{Tetrahedra}. As $T$ is increased, excitations above this ground state will start to appear, leading to a transition into the Coulomb phase described in Section~\ref{SecCoulombPhase}. For $T, h \ll J$, however, a single flipped spin is not a low-energy excitation: it breaks the ice-rule constraint, and hence costs an energy on the order of $J$.

In order to satisfy the ice rules, every tetrahedron must have the same number of flipped spins at the top as at the bottom; an example is shown on the right-hand side of Figure~\ref{Tetrahedra}. This in turn implies that every $(100)$ plane of the lattice should have the same number of flipped spins, and the lowest-energy excitation consistent with the ice rules is a string of flipped spins spanning the system in the $[100]$ direction. Such a string has a Zeeman energy of $2h/\sqrt{3}$ per unit of its length.

An isolated string can be modelled as a random walk through the lattice in the $[100]$ direction, with two choices at every step. There is therefore an entropy of $\ln 2$ per unit length of the string, and the free energy $F$ in the presence of a single string is\cite{Jaubert}
\beq{FreeEnergy}
F = L_z\left(\frac{2}{\sqrt{3}}h - T \ln 2\right)\punc{,}
\eeq
where $L_z$ is the length of the system in the $z$-direction. As a result, for $L_z \rightarrow \infty$, strings are absent below a critical temperature $T\sub{K} = 2h/\sqrt{3}\ln 2$. For $T > T\sub{K}$, however, the entropic gain from introducing these string excitations into the system outweighs their Zeeman energy cost, giving a phase transition into the Coulomb phase.

As noted previously,\cite{Jaubert} this transition has a strongly asymmetric character and is an example of a Kasteleyn transition.\cite{Kasteleyn,Bhattacharjee} On the ordered side of the transition, fluctuations are completely suppressed and the transition appears to be of first order, while there are fluctuations in the Coulomb phase and a divergent correlation length as the transition approaches, as for a continuous transition.

The entropic argument as presented so far is only valid in the limit of infinite $J$, where the ice rules are strictly enforced on every tetrahedron. Finite $J/T$ allows thermal excitations that break the ice rules at a small density of tetrahedra, leading to a characteristic string length $L\sub{c} \sim \ee^{\alpha J / T}$, where $\alpha$ is a numerical constant. For $L\sub{c} < L_z$, this characteristic length replaces $L_z$ in \refeq{FreeEnergy}. The effect of finite $J/T$ is therefore similar to the effect of a finite system size, replacing the phase transition by a crossover, whose sharpness increases with $J/T$.

In the infinite-$J$ limit, the description in terms of strings remains valid for all $h$: any state obeying the ice rules can be constructed by starting from the $h \rightarrow \infty$ configuration and flipping all spins along a set of strings, with each spanning the system in the $[100]$ direction. In other words, there is a one-to-one mapping from spin configurations obeying the ice rules to string configurations.\cite{Jaubert}

\section{Boson world-lines}
\label{SecBosons}

In order to understand the phase transition in detail, it is convenient to treat the strings as world-lines for bosons moving in a two-dimensional space, with the $[100]$ direction taken as imaginary time. The classical partition function is then interpreted as the partition function for a quantum problem, with the length of the system in the imaginary-time direction corresponding to the inverse temperature. In the thermodynamic limit we therefore have a two-dimensional quantum problem at zero temperature.

Since the strings span the system in the $[100]$ direction, the boson number is conserved and there are no closed loops, which would correspond to particle--antiparticle pairs. The mapping therefore provides a way to view the Kasteleyn transition as the zero-temperature quantum phase transition between the vacuum and a Bose-Einstein condensate, as the chemical potential for bosons $\mu$ increases through some critical value $\mu _0$.

The vacuum of bosons then corresponds to the saturated state of the magnet at high field or low temperature. The lowest excitations in the quantum problem are single particles, which correspond to isolated strings. These have finite energy above the ground state and so the correlation functions are not long ranged. In fact, since the classical observables map to operators that are diagonal in the boson-number basis, their connected correlation functions are strictly zero in this phase.

In the other phase, with a Bose condensate, there is a Goldstone mode corresponding to the broken phase-rotation symmetry, leading to power-law correlation functions for the spins. We will show in Section~\ref{SecContinuumTheory} that the correlations have the expected dipolar (three-dimensional) spatial dependence (a simplified treatment has been outlined previously\cite{Jaubert}). These long-range correlations remain as the field is decreased to zero, when they agree with the predictions of Section~\ref{SecCoulombPhase}.

In Section~\ref{ODLRO}, we will provide further evidence for the identification of the Coulomb phase with the condensed phase of the bosons, by showing that the off-diagonal long-range order in this phase can be interpreted in terms of deconfinement in the Coulomb phase.

As conjectured earlier,\cite{Jaubert} and as we will demonstrate explicitly in Section \ref{SecCriticalTheory}, the condensation phase transition has dynamical critical exponent $z=2$, and so our $(2+1)$-dimensional system is at its upper critical dimension. Thermodynamic results for the boson problem\cite{Schick,FisherHohenberg,Posazhennikova} can then be carried over directly to the spin system.\cite{Jaubert}

As an aside, we note that the effect of finite $J/T$ can also be straightforwardly understood using this picture. Tetrahedra that break the ice rules, which appear with nonzero density away from the infinite-$J$ limit, become the ends of strings. In the quantum Hamiltonian, these are generated by source terms for the boson operator, which break the phase rotation symmetry and eliminate the phase transition. As noted previously, a large but finite $J/T$ therefore has the effect of replacing the phase transition with a sharp crossover.

\subsection{Effective Hamiltonian}
\label{EffHam}

We will now present a more detailed analysis of the mapping from the problem defined in terms of spins to a quantum problem of bosons, which will allow the calculation of correlation functions, for both zero and nonzero field. The mapping essentially follows the standard procedure of using a transfer matrix to connect the degrees of freedom in one layer of the system to those in the next, followed by interpretation of the transfer matrix as the exponential of a quantum Hamiltonian. However, care must be taken to account properly for the pyrochlore structure.

We first divide the lattice sites by their $z$ coordinates, which we take to be measured along the $[100]$ direction. Each spin in the pyrochlore lattice has neighbours only in the same plane and in those immediately above and below. The partition function can therefore be written in terms of a transfer matrix whose rows and columns are labelled by the possible configurations of a given plane of the lattice.

\begin{figure}
\putinscaledfigure{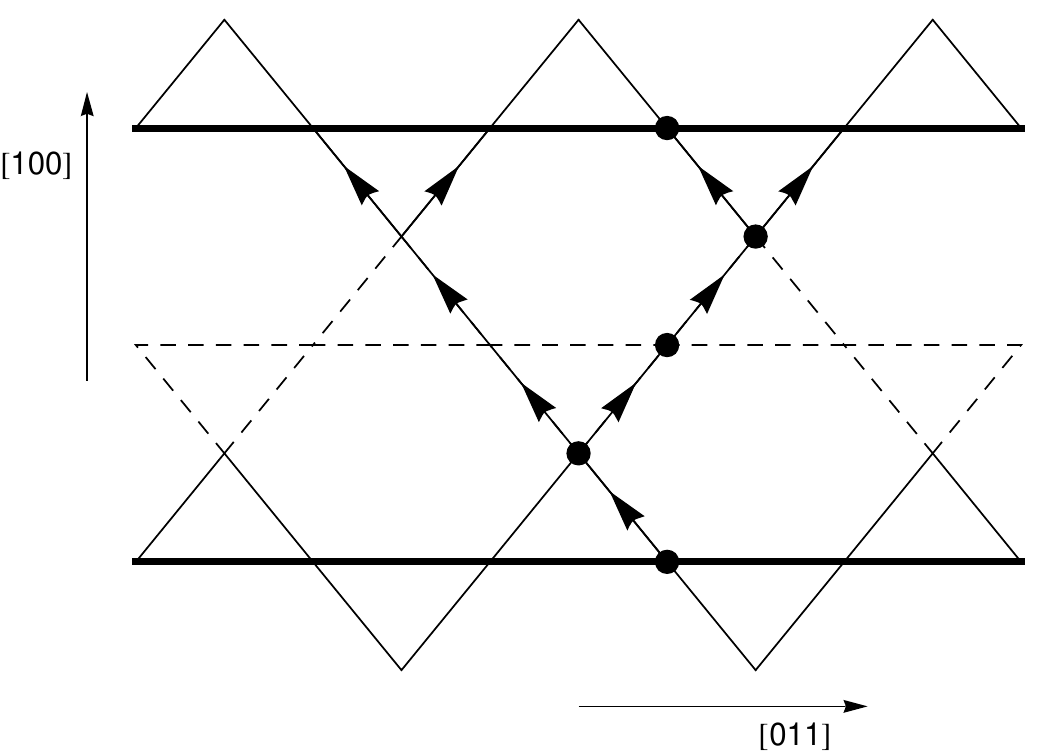}
\caption{\label{Pyrochlore011Projection}Projection of the pyrochlore lattice onto the $(01\bar{1})$ plane. The sites of the pyrochlore lattice are at the vertices. The tetrahedra appear as triangles, and the dashed lines indicate tetrahedra that lie out of the plane. The two thick horizontal lines correspond to the thick lines shown in Figure~\ref{PyrochloreWith2DPlanes} and represent one choice for the $(100)$ planes that define the quantum problem. The distance between them in the $[100]$ direction is one step in imaginary time, $\Delta\tau$. The points show a possible path for a string between these two planes, and the arrows indicate all possible locations for the end of the string, given its starting position. There are more points to the left of the starting location than to the right, and so a boson at this point has a preference for leftward hopping.}
\end{figure}

The structure of pyrochlore is such that the positions of the sites in neighbouring $(100)$ planes do not coincide, and it is in fact necessary to translate by four layers before the lattice structure in a plane repeats. This is illustrated in Figure~\ref{Pyrochlore011Projection}, which shows the projection of the pyrochlore lattice onto the $(01\bar{1})$ plane. The thick horizontal lines show the positions of two nearest planes with the same site positions, which are separated by three planes with sites in different positions. We therefore define a new transfer matrix corresponding to the product of four elementary transfer matrices, which relates two such planes with the same lattice structure.

This transfer matrix $\T$ can then be interpreted in the standard way as the exponential of the Hamiltonian matrix $\quH$, by writing $\T = \ee^{-\quH \Delta \tau}$, where $\Delta\tau$ is a step in imaginary time. The trace over a product of transfer matrices corresponds to the standard sum over histories giving the partition function for the two-dimensional quantum problem.

While it is possible to construct the matrix $\T$ explicitly by considering all arrangements of the spins in the three intermediate layers, and therefore in principle possible to find $\quH$ exactly, we do not expect this to be necessary for our purposes. Instead, we will attempt to understand the general form of the Hamiltonian based on the symmetries of the problem, while restricting to the simplest version that is necessary to describe the large-scale behaviour correctly.

First note that, since it describes a classical statistical model, the matrix elements of $\T$ are required to be real and nonnegative. We can therefore restrict to the case where $\quH$ is also a real matrix. However, as is often the case for classical statistical problems, the transfer matrix, and hence Hamiltonian, are not required to be hermitian. In fact, we will find (see Sections~\ref{SecCondensedPhase} and \ref{LargeSCorrFunc}) that the sublattice structure of the correlation functions (see Section~\ref{SecCoulombPhase}) is crucially dependent on the nonhermitian terms in the Hamiltonian, and so we will pay particular attention to such terms.

The fact that $\T$ has all positive elements implies that its eigenvalue with largest real part---corresponding to the ground-state energy of $\quH$---is unique (within each boson-number sector), is positive, and has an eigenvector with all positive weights. This can be interpreted as meaning that the bosonic Hamiltonian $\quH$ is unfrustrated, which is consistent with our assumption of a superfluid ground state.

As described above, we will represent the state of a given layer by comparing to a reference configuration in which all spins have a positive projection on the $+z$ axis. This configuration will be referred to as the vacuum of bosons, while flipped spins are mapped to the presence of a boson. The bosons therefore have on-site hard-core repulsion by construction, and as noted above, the ice rules enforce conservation of boson number.

We define the bosonic annihilation and creation operators $b_i$ and $b_i^\dagger$ for sites $i$ in the two-dimensional lattice. The hard-core constraint on the bosons is implemented by the operator identities $b_i b_i = b_i^\dagger b_i^\dagger = 0$, so that the number operator $n_i = b_i^\dagger b_i$ is restricted to the values $0$ and $1$. (Operators for different sites commute.)

We will divide the Hamiltonian $\quH$ into potential terms $V$ and kinetic terms $K$, by writing
\beq{HVK}
\quH = V + K\punc{,}
\eeq
and address these two parts in turn.

The simplest contribution to $V$ is a chemical potential term $V_0$ giving the energy cost of adding a boson world-line to the system. In the presence of an applied magnetic field along the $+z$ axis, every flipped spin has Zeeman energy cost $2h\sub{eff}$, so that we can write
\beq{ChemicalPotential}
V_0 = -\mu \sum_i \left(n_i - \frac{1}{2}\right)\punc{,}
\eeq
where $\mu \propto -h$. The classical model is symmetric under $h \rightarrow -h$ along with an inversion of all the spins. After mapping to the quantum problem, this takes $\mu \rightarrow -\mu$ and exchanges particles and holes; the constant ($-\frac{1}{2}$) has been chosen to make this symmetry explicit. A vanishing applied field corresponds to half filling in the boson representation, and at this point the model is particle--hole symmetric.

Interactions between the bosons are dominated by the hard-core repulsion, but there will also be off-site interactions that contribute to $V$. These terms are necessarily hermitian and we will not consider them further.

The kinetic terms $K$ will include both quadratic hopping and more complicated correlated hopping terms, and, as we have observed above, are not required to be hermitian. For a single boson, the physical origin of the nonhermitian terms is clear from the lattice structure. As shown in Figure~\ref{Pyrochlore011Projection}, a string starting on a given site has a tendency to hop in one direction in preference to the opposite direction, and it is this directed hopping that requires the Hamiltonian to be nonhermitian.

Note, however, that a hole surrounded by particles hops preferentially in the same direction as an isolated particle, as can be checked by exchanging particles and holes in Figure~\ref{Pyrochlore011Projection}. As a result, the quadratic hopping term,
\beq{eqK1}
K_1 = -\sum_{ij} t_{ij} b^\dagger_i b_j\punc{,}
\eeq
is required to be symmetric under $b\nd _i \leftrightarrow b^\dagger _i$, so that $t_{ij} = t_{ji}$ and $K_1$ is hermitian. The same conclusion follows from the requirement of particle--hole symmetry when $h=0$.

We must therefore include higher-order terms in order to describe the directed hopping, and so we seek correlated hopping terms that are particle--hole symmetric. The simplest such term is
\beq{eqK2}
K_2 = -\sum_{ij\ell} w_{ij\ell} (n_\ell - {\textstyle\frac{1}{2}})b^\dagger_i b\nd _j\punc{,}
\eeq
where $w_{ij\ell} = -w_{ji\ell}$ and so $K_2 = -K_2^\dagger$. Note that near the transition, where the density is small, $K_2$ reduces to a quadratic hopping term of the form of $K_1$, but with effective directed hopping $t^{\mathrm{eff}}_{ij} = -\frac{1}{2}\sum _\ell w_{ij\ell}$.

We therefore take $V = V_0$ and $K = K_1 + K_2$. The coefficients $t_{ij}$ and $w_{ij\ell}$ are required to be real and chosen so that, as noted above, the hopping is unfrustrated.

\subsection{Symmetries}
\label{SecSymmetries}

In constraining the general form of terms that can appear in the effective Hamiltonian and the corresponding continuum action, it is important to consider symmetries inherited from the original problem defined on the pyrochlore lattice. The choice of $z$ as the imaginary time axis and of the particular $(100)$ planes as described in Section~\ref{EffHam} reduces the symmetry group somewhat.

\begin{figure}
\putinscaledfigure{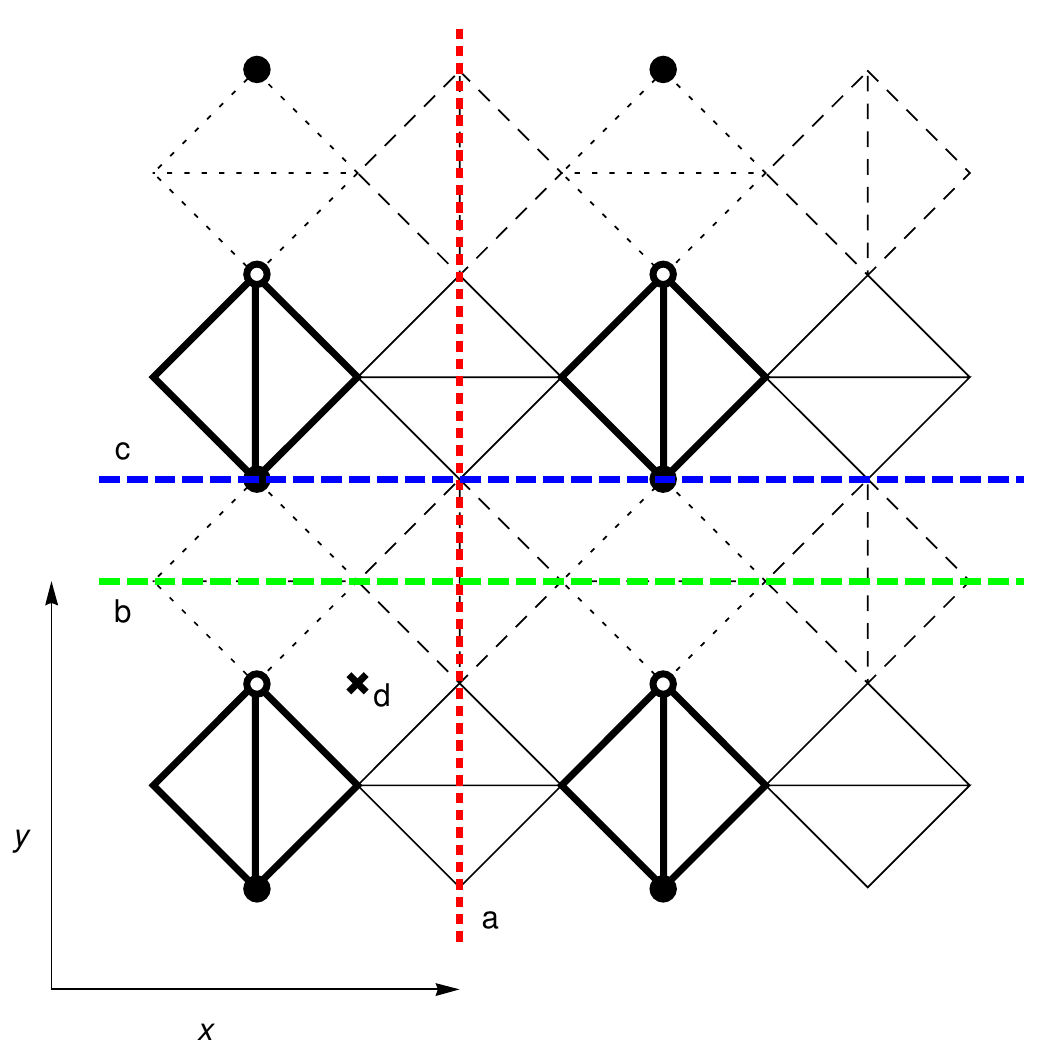}
\caption{\label{Pyrochlore100Projection}Projection of the pyrochlore lattice onto the $(100)$ plane, with the $x$- and $y$-axes corresponding to the $[01\bar{1}]$ and $[011]$ directions respectively. Each tetrahedron is shown as a square with a line joining the two uppermost points. The weight of the lines indicates the height of a given tetrahedron in the $[100]$ direction, with thick lines highest, followed by thin, dashed, and dotted lines respectively. The strucure is periodic, so that another tetrahedron of the same orientation lies four layers below (and above) each tetrahedron that is shown. The circles indicate sites in a particular $(100)$ plane, which, after the mapping to the quantum problem, become the sites for the bosons. Filled and empty circles correspond to the black and white points shown in Figure~\ref{PyrochloreWith2DPlanes} and denote the two sublattices, which are distinguished by the structure of the surrounding lattice. The thick red and green lines (labelled a and b) show planes of reflection symmetry, while there is a symmetry under rotation by $180^\circ$ about the blue line (labelled c). Rotation by $\frac{\pi}{2}$ about the cross (labelled d) and translation in the $[100]$ direction maps sites in one $(100)$ plane into those in the plane immediately above.}
\end{figure}

This is illustrated in Figure~\ref{Pyrochlore100Projection}, which shows the pyrochlore lattice projected onto the $(100)$ plane. It can be seen immediately from the arrangement of tetrahedra that nearest neighbours in the $x$-direction are equivalent, whereas those in the $y$-direction are not. The lattice can therefore be represented as two interpenetrating square lattices, indicated by the filled and empty circles. We define the elementary translation operators $\xTrans$ and $\yTrans$ which take sites into their nearest equivalent neighbours in the $x$- and $y$-directions respectively, and which are symmetries of the Hamiltonian. (Note that $\xTrans$ and $\yTrans$ in fact translate by the same absolute distance, although $\xTrans$ connects neighbours, while $\yTrans$ connects next-neighbours in the $y$-direction.) We also define the operation $\zTrans$, a translation by four layers in the $z$-direction, which corresponds to time-translation symmetry in the quantum problem.

One can in addition see that there is a reflection symmetry, which we label $\xReflect$, about the red line labelled a, which takes $x\rightarrow -x$ but does not exchange sublattices. There is similarly a reflection symmetry $\yReflect$ about the green line labelled b, taking $y\rightarrow -y$, but in this case the symmetry exchanges the two sublattices.

Besides $\xReflect$ and $\yReflect$, there is a third reflection symmetry, which involves reflection in the blue line labelled c, and also takes $y\rightarrow -y$. This does not affect the sublattices, but changes the height of the tetrahedra: the tetrahedra drawn with thick lines in the figure are exchanged with those drawn with dotted lines and those with thin lines are exchanged with those with dashed lines. This amounts to a reflection in the $x$-$y$ plane (i.e., $z \rightarrow -z$), which taken together with the reflection about the $x$-direction constitutes a third reflection symmetry, $\yzReflect$. In three dimensions, $\yzReflect$ is a rotation by $180^\circ$ about the $x$-direction, while, after mapping to the quantum problem defined in two dimensions, it becomes a $y$-reflection accompanied by a time-reversal operation.

Finally, consider once more the elementary transfer matrix linking one $(100)$ plane to that immediately above it. As explained above, this matrix does not act within the quantum Hilbert space, since it relates sites on two different (two-dimensional) lattice structures. Instead, we define the operator $\T_{1/4}$, given by this elementary `evolution' operator followed by a rotation by $\frac{\pi}{2}$ about the cross labelled d in Figure~\ref{Pyrochlore100Projection}. This composite transformation returns the sites to their original locations and hence acts within the Hilbert space. It is also clear that it commutes with the Hamiltonian, since the original choice of a particular $(100)$ plane was arbitrary.

As for the full transfer matrix $\T$, we will not write down the precise form of the operator $\T_{1/4}$, and we will not attempt to find the general constraints it implies for the Hamiltonian. Instead, we will merely note that, in a coarse-grained description, it has a particularly simple effect on operators that do not distinguish the two sublattices, giving only a $\frac{\pi}{2}$ rotation in the $x$-$y$ plane.

\subsection{Off-diagonal long-range order and deconfinement}
\label{ODLRO}

As we have explained above, we identify the Coulomb phase of the classical statistical model with the condensed phase of the quantum bosons. In Section~\ref{SecCondensedPhase}, we will show that the long-range correlations in the quantum model reproduce those of the classical model, supporting this identification. Here, we present a qualitative argument based on the concept of off-diagonal long-range order (ODLRO).

First, note that while the quantum model has $\langle b_i \rangle \neq 0$ in the condensed phase if total boson number is allowed to fluctuate, there is no equivalent order parameter in the Coulomb phase. Under the mapping we use, boson number is strictly conserved and a nonzero order parameter $\langle b_i \rangle$ is impossible. Spontaneous symmetry breaking in the quantum problem is therefore hidden in the statistical-mechanics equivalent. We instead consider ODLRO: in the condensed phase, the correlation function $\langle b_i^\dagger b_j\nd \rangle$ has a nonzero limit for large separation of the sites $i$ and $j$, given by the square modulus of the order parameter.

Translating this quantum expectation value into the language of the statistical problem gives $\langle b_i^\dagger b_j \rangle \sim {\Z_{ij}}/{\Z}$. Here, with $\beta$ the quantum inverse temperature (to be taken to infinity), $\Z = \Tr \T ^\beta$ is the partition function, and $\Z_{ij}$ is given by
\beq{Zij}
\Z_{ij} = \Tr \T^{\beta} b_i^\dagger b_j\nd\punc{.}
\eeq
This can be interpreted as the partition function, calculated in the presence of one string (boson world-line) that ends at the point $j$ on a certain arbitrary $(100)$ plane and another that begins at the point $i$ on the same plane.

This is most simply interpreted by returning to the representation in terms of solenoidal fields described in Section~\ref{SecCoulombPhase}. A string ending at a given point corresponds to a tetrahedron on which the ice rules are broken, so that the coarse-grained field $\Bv(\rv)$ has nonzero divergence. We therefore have two monopoles of opposite `charge' separated by a large distance.

In a confining phase of the field $\Bv$, the energy increases linearly with separation, so that ${\Z_{ij}}/{\Z}$ is exponentially suppressed for large separations. In the Coulomb phase, however, monopoles are deconfined, and ${\Z_{ij}}/{\Z}$ approaches a constant, given by the monopole fugacity, for large separation.

The ODLRO in the quantum superfluid is therefore equivalent to monopole deconfinement,\cite{Castelnovo} in agreement with our identification of the Coulomb and superfluid phases.

\section{Continuum theory}
\label{SecContinuumTheory}

In this section, we show that the long-range form of the correlation functions in spin ice, with and without an applied field, can be found using an effective continuum theory for the hard-core bosons. We also find a critical theory for the Kasteleyn transition, and show that, although the directed hopping is necessary for the appropriate long-range correlation functions, it does not affect the critical behavior. In Section \ref{SecLargeS}, we show that the same long-range behavior can be found directly from the microscopic quantum Hamiltonian, by using a large-$S$ expansion.

Our approach here is to start with the Hamiltonian $\quH$ and write down the most general continuum action consistent with the symmetries of the hard-core boson problem, described in Section~\ref{SecSymmetries}. In the Appendix, we sketch an approach that would in principle allow such a continuum action to be derived directly. The action is expressed in terms of bosonic fields corresponding to the hard-core boson operator $b$, in the limit where the spatial coordinates $x$ and $y$ and the imaginary time $\tau$ are taken as continuous. Since it is important to preserve the two-sublattice structure, we define (c-number) fields $\psi_\sublattice{a}$ and $\psi_\sublattice{b}$ corresponding to the two sublattices.

Besides the spatial symmetries, there is a phase-rotation symmetry arising from the conservation of the boson number, which leaves the action invariant under the simultaneous transformations $\psi_\sublattice{a}\rightarrow\psi_\sublattice{a}\ee^{\ii\theta}$ and $\psi_\sublattice{b}\rightarrow\psi_\sublattice{b}\ee^{\ii\theta}$. Note that only the total number of bosons on both sublattices is conserved, and so terms such as $\psi_\sublattice{a}^*\psi_\sublattice{b}\ns$ are allowed by this symmetry.

We now consider each of the symmetries defined in Section~\ref{SecSymmetries} in turn. The translations $\xTrans$, $\yTrans$, and $\zTrans$ shift $x$, $y$, and $\tau$ (infinitesimally, in the continuum limit) and so terms with explicit dependence on the coordinates are forbidden.

As for the reflection symmetries, $\xReflect$ simply takes $x\rightarrow -x$, while $\yReflect$ takes $y\rightarrow -y$, simultaneously exchanging the two sublattices. The derivative operator $\partial _x \equiv \partial/\partial x$ can therefore only appear in the combination $\partial _x^2$, but a single $y$-derivative can appear in a combination such as $(\psi_\sublattice{a} - \psi_\sublattice{b})\partial _y$.

In the continuum theory for standard lattice boson models,\cite{Subir} one also generically has symmetry under time reversal accompanied by particle--hole exchange: $\partial _\tau \rightarrow -\partial _\tau$, $\psi\rightarrow\psi^*$; and, at points where the theory is particle--hole symmetric, such as half-filling for hard-core bosons, the two transformations are separately symmetries. In our problem, the nonhermitian nature of the Hamiltonian means that this is no longer the case. Instead, as we show in the Appendix, the reflection operation $\yzReflect$ leads to a symmetry under $\partial _y\rightarrow -\partial _y$, $\partial _\tau\rightarrow -\partial _\tau$, $\psi_{\sublattice{a},\sublattice{b}}\ns\rightarrow\psi_{\sublattice{a},\sublattice{b}}^*$. At half-filling, the particle--hole transformation $\psi_{\sublattice{a},\sublattice{b}}\ns\rightarrow\psi_{\sublattice{a},\sublattice{b}}^*$ is again a separate symmetry, which we label $\phInvert$.

The most general Lagrangian density that is consistent with these symmetry constraints, up to quadratic order in the fields and second derivatives (and omitting terms that are total derivatives), can be written
\begin{widetext}
\begin{multline}
\label{Lagrangian}
\Lag = r_1(\psi_\sublattice{a}^*\psi_\sublattice{a}\ns+\psi_\sublattice{b}^*\psi_\sublattice{b}\ns) + r_2(\psi_\sublattice{a}^*\psi_\sublattice{b}\ns+\psi_\sublattice{b}^*\psi_\sublattice{a}\ns)
+ \ii\kappa _\tau (\psi_\sublattice{a}^*\partial _\tau \psi_\sublattice{a}\ns + \psi_\sublattice{b}^*\partial _\tau \psi_\sublattice{b}\ns)
+ \ii\kappa _\tau ' (\psi_\sublattice{a}^*\partial _\tau \psi_\sublattice{b}\ns + \psi_\sublattice{b}^*\partial _\tau \psi_\sublattice{a}\ns)\\
+ \ii\kappa _y (\psi_\sublattice{a}^*\partial _y \psi_\sublattice{a}\ns - \psi_\sublattice{b}^*\partial _y \psi_\sublattice{b}\ns)
+\lambda _\mu \left[(\partial _\mu \psi_\sublattice{a}^*)(\partial _\mu \psi_\sublattice{a}\ns) + (\partial _\mu \psi_\sublattice{b}^*)(\partial _\mu \psi_\sublattice{b}\ns) \right]
+ \lambda _\mu ' \left[(\partial _\mu \psi_\sublattice{a}^*)(\partial _\mu \psi_\sublattice{b}\ns) + (\partial _\mu \psi_\sublattice{b}^*)(\partial _\mu \psi_\sublattice{a}\ns)\right] + \;\cdots\punc{,}
\end{multline}
\end{widetext}
where a sum over $\mu\in\{x,y,\tau\}$ is implied. At half-filling, particle--hole symmetry implies $\kappa _\tau = \kappa _\tau ' = \kappa _y = 0$. (Note that we so far have made no use of the symmetry $\T_{1/4}$, which has a nontrivial effect on the sublattice labels $\sublattice{a}$ and $\sublattice{b}$.)

To evaluate the correlation functions, it is also necessary to find expressions for the boson number operators in terms of the fields $\psi_\sublattice{a}$ and $\psi_\sublattice{b}$, and symmetry considerations can again be used. Defining fields $n_\sublattice{a}$ and $n_\sublattice{b}$ representing the number fluctuations on the two sublattices, we have $n_\sublattice{a} \leftrightarrow n_\sublattice{b}$ under $\yReflect$ and $n_{\sublattice{a},\sublattice{b}}\rightarrow -n_{\sublattice{a},\sublattice{b}}$ under $\phInvert$; they are unaffected by all other symmetries. (Note that these are the number fluctuations; the actual number operators also contain constant terms, which will cancel when calculating connected correlation functions.) These symmetries are not, however, sufficient to determine the number operators completely. We find, for instance, that $n_\sublattice{a}+n_\sublattice{b}$ is given by a linear combination of $\ii(\psi_\sublattice{a}^*\partial _\tau \psi_\sublattice{a}\ns + \psi_\sublattice{b}^*\partial _\tau \psi_\sublattice{b}\ns)$, $\ii(\psi_\sublattice{a}^*\partial _\tau \psi_\sublattice{b}\ns + \psi_\sublattice{b}^*\partial _\tau \psi_\sublattice{a}\ns)$, and $\ii(\psi_\sublattice{a}^*\partial _y \psi_\sublattice{a}\ns - \psi_\sublattice{b}^*\partial _y \psi_\sublattice{b}\ns)$, where the relative coefficients cannot be determined by this method.

\subsection{Critical theory}
\label{SecCriticalTheory}

Using the Lagrangian density given in \refeq{Lagrangian}, we can address the behaviour at the Kasteleyn transition from the vacuum to a phase with a Bose-Einstein condensate. This necessarily occurs away from half filling, where there is no particle--hole symmetry (although an identical transition, related by a particle--hole transformation, occurs at unit filling). At the transition, both $\psi_\sublattice{a}$ and $\psi_\sublattice{b}$ acquire nonzero expectation values, breaking phase-rotation symmetry.

It is convenient to transform from $\psi_\sublattice{a}$ and $\psi_\sublattice{b}$ to a new basis that diagonalizes the constant, quadratic part of $\Lag$. Due to the symmetry $\yReflect$, which exchanges the two sublattices, the appropriate fields are $\psi_{\pm} = \psi_\sublattice{a}\pm\psi_\sublattice{b}$, and the constant part of the Lagrangian density then becomes
\beq{ConstLagrangian}
\Lag\sub{const} = r_+ \psi_+^*\psi_+\ns + r_- \psi_-^*\psi_-\ns + \;\cdots\punc{.}
\eeq
The difference of the two constants $r_+$ and $r_-$ is given by $r_2$ in \refeq{Lagrangian}, and will generically be nonzero. Both $r_+$ and $r_-$ will be positive in the vacuum phase and will grow smaller as the transition is approached. At the transition, one of the two (strictly, its renormalized value) changes sign, and the corresponding field $\psi_{\pm}$ acquires a nonzero expectation value. As a result, $\langle\psi_\sublattice{a}\rangle$ and $\langle\psi_\sublattice{b}\rangle$ both become nonzero, and with the same magnitude.

The theory can then be rewritten in terms of the critical field, $\psi _+$ say, while the other ($\psi _-$), which still has a `mass' term is integrated out. To determine the form of this critical theory, consider the quadratic terms with single derivatives. Rewriting the term in $\Lag$ with one $\partial _y$ derivative in terms of $\psi _{\pm}$ gives $\psi_\sublattice{a}^*\partial _y \psi_\sublattice{a}\ns - \psi_\sublattice{b}^*\partial _y \psi_\sublattice{b}\ns \sim \psi_+^*\partial _y \psi_-\ns + \psi_-^*\partial _y \psi_+\ns$. Integrating out the field $\psi_-$ leaves only a term of the form $(\partial _y \psi_+^*)(\partial _y \psi_+\ns)$. On the other hand (as long as $\kappa _\tau + \kappa _\tau ' \neq 0$), there will be a term $\psi _+^*\partial _\tau \psi _+\ns$ in the Lagrangian density, which remains after $\psi_-$ is integrated out.

Furthermore, since the field $\psi _+$ has no dependence on the sublattice structure, we make use of the symmetry $\T_{1/4}$ to require invariance under a $\frac{\pi}{2}$ rotation in the $x$-$y$ plane. (The critical theory therefore has an emergent symmetry under this operation, and in fact under continuous rotations in the $x$-$y$ plane.) The critical theory, which is written only in terms of $\psi _+$, then takes the form
\beq{CriticalTheory}
\Lag\sub{critical} = r_+ \psi_+^*\psi_+\ns + \ii \psi _+^*\partial _\tau \psi _+\ns + \lambda^+ (\partial _j \psi_+^*)(\partial _j \psi_+\ns) + \;\cdots\punc{,}
\eeq
where a sum over $j\in \{x,y\}$ is implied: the second time derivative has been dropped. This therefore describes the standard vacuum transition of bosons, with dynamical critical exponent $z=2$, as conjectured previously.\cite{Jaubert}

\subsection{Condensed phase}
\label{SecCondensedPhase}

We can also use the continuum theory of \refeq{Lagrangian} to calculate correlation functions within the condensed phase. Once the fields $\psi_\sublattice{a}$ and $\psi_\sublattice{b}$ acquire nonzero expectation values, they are most conveniently described in terms of phase and amplitude modes, by writing
\beq{PhaseAmplitude}
\psi_\sublattice{a} = \sqrt{\rho _0 + \delta\rho_\sublattice{a}}\ee^{\ii(\phi + \theta)}
\quad\text{and}\quad
\psi_\sublattice{b} = \sqrt{\rho _0 + \delta\rho_\sublattice{b}}\ee^{\ii(\phi - \theta)}\punc{,}
\eeq
where $\rho _0$ is the (real) condensate density, and $\delta\rho_{\sublattice{a},\sublattice{b}}$, $\phi$, and $\theta$ are real fields.

The amplitude modes, $\delta\rho_\sublattice{a}$ and $\delta\rho_\sublattice{b}$, and the relative phase mode, $\theta$, correspond to gapped excitations, while the overall phase $\phi$ is a Goldstone mode, resulting from the broken phase-rotation symmetry of the condensate. Only this gapless mode can contribute to the long-range correlation functions, and we therefore integrate out the remaining gapped modes. While this process can be carried out explicitly, it is clear from symmetry considerations that the general form for the resulting Lagrangian is simply $\Lag_\phi = (\partial _\tau\phi)^2 + c^2 (\partial _j \phi)^2 + \cdots$,
where a sum over $j \in \{x,y\}$ is again implied. The constant $c$, giving the speed of sound in the condensate, cannot be determined by symmetry.

In terms of the field $\phi$, the number operators take a simple form: symmetry requires that we have
\beq{BosonNumbers2}
n_\sublattice{a} \sim (\partial _\tau + v\partial _y) \phi
\qquad\text{and}\qquad
n_\sublattice{b} \sim (\partial _\tau - v\partial _y) \phi\punc{,}
\eeq
but does not fix the magnitude or sign of $v$.

Taking the Fourier transform therefore gives for the correlation functions
\begin{align}
\label{nnCorrelation1}
\langle n_\sublattice{a} n_\sublattice{a} \rangle &\sim \frac{({\omega + v k_y)}^2}{\omega^2 + c^2 (k_x^2 + k_y^2)}\\
\langle n_\sublattice{b} n_\sublattice{b} \rangle &\sim \frac{({\omega - v k_y)}^2}{\omega^2 + c^2 (k_x^2 + k_y^2)}\\
\langle n_\sublattice{a} n_\sublattice{b} \rangle &\sim \frac{\omega^2 - v^2 k_y^2}{\omega^2 + c^2 (k_x^2 + k_y^2)}
\label{nnCorrelation3}\punc{.}
\end{align}
After the identification of $\omega$ with $k_{[100]}$ and $k_y$ with $k_{[011]}$, these become exactly equivalent to Eqs.~(\ref{aaCorrFunc}) and (\ref{abCorrFunc}), derived in Section~\ref{SecCoulombPhase} using the mapping to a solenoidal field. Note that Eqs.~(\ref{aaCorrFunc}) and (\ref{abCorrFunc}) apply only in the absence of an applied magnetic field, whereas Eqs.~(\ref{nnCorrelation1}) to (\ref{nnCorrelation3}) are valid throughout the Coulomb phase, with the dependence on $h$ appearing in the constants $c$ and $v$. These cannot be determined by this method, however, since the quantum mapping explicitly reduces the symmetry of the three-dimensional problem by picking out the $z$-direction.

As the vacuum transition is approached from the condensed phase, the speed of sound decreases continuously towards zero, and standard results for the boson problem\cite{Schick} give $c \sim \sqrt{\mu - \mu _0}$. This leads in spin ice to anisotropy in the correlation functions between the $z$-direction and the $x$- and $y$-directions, with relative scale $c \sim \sqrt{h\sub{K} - h}$, where $h\sub{K}$ is the critical field at the Kasteleyn transition.

\section{Large-$S$ calculation of correlation functions}
\label{SecLargeS}

We now present an alternative calculation of the spin--spin correlation function in the Coulomb phase, starting from the Hamiltonian in the bosonic representation. Our approach is to rewrite the hard-core bosons in terms of $S=\frac{1}{2}$ spins and then use a spin-wave approximation based on a large-$S$ expansion.

\subsection{Holstein-Primakoff expansion}

We begin with the effective Hamiltonian described in Section~\ref{EffHam},
\beq{FullHam}
\begin{split}
\quH &= V_0 + K_1 + K_2\\
&= -\mu \sum_i \left(n_i - \frac{1}{2}\right)
-\sum_{ij} t_{ij} b^\dagger_i b_j\\
&\qquad\qquad\qquad\qquad\qquad-\sum_{ij\ell} w_{ij\ell} (n_\ell - {\textstyle\frac{1}{2}})b^\dagger_i b_j\punc{,}
\end{split}
\eeq
where the coefficients $\mu$, $t_{ij}$ and $w_{ij\ell}$ are real and obey $t_{ij} = t_{ji}$ and $w_{ij\ell} = -w_{ji\ell}$. We first map the hard-core bosons into $S=\frac{1}{2}$ spins by identifying $n_i \equiv b_i^\dagger b_i\nd = \frac{1}{2} - \Sigma_i^z$, $b_i^\dagger = \Sigma_i^-$, and $b_i = \Sigma_i^+$, where $\Sigma_i^{\pm} = \Sigma_i^x \pm \ii \Sigma_i^y$. (We use the symbol $\Sigma$ rather than $S$ to avoid confusion with the classical spins of the original spin ice Hamiltonian, to which these quantum spins are not simply related. In particular, the labels $x$, $y$, and $z$ do not correspond to the directions of the classical spins or the crystal structure.) In terms of spins, the chemical potential becomes an applied field along the $z$ direction, while the kinetic terms produce couplings between the $x$ and $y$ components of nearby spins.

We now extend this model to $S \ge \frac{1}{2}$, so that $\boldsymbol{\Sigma}_i$ become spin-$S$ operators for general $S$. For large $S$, the Holstein-Primakoff transformation can be used to expand around a classical ground state. Such states have all spins parallel and at an angle $\theta$ to the field, and we choose the state where they lie in the $x$-$z$ plane. Rotating to new axes aligned with the classical spins, we write
\begin{align}
\label{HPstart}
\Sigma_i^z& = \tilde\Sigma_i^z \cos\theta - \tilde\Sigma_i^x \sin\theta\\
\Sigma_i^x& = \tilde\Sigma_i^z \sin\theta + \tilde\Sigma_i^x \cos\theta\\
\Sigma_i^y& = \tilde\Sigma_i^y\punc{.}
\end{align}
The Holstein-Primakoff transformation is given by
\begin{align}
\tilde\Sigma_i^+ &= \sqrt{2S} \sqrt{1-\frac{a_i^\dagger a_i\nd}{2S}} a_i\\
\tilde\Sigma_i^- &= \sqrt{2S} a_i^\dagger \sqrt{1-\frac{a_i^\dagger a_i\nd}{2S}}\\
\tilde\Sigma_i^z &= S - a_i^\dagger a_i\punc{,}
\label{HPend}
\end{align}
where $\tilde\Sigma_i^\pm = \tilde\Sigma_i^x \pm \ii \tilde\Sigma_i^y$, and $a_i$ and $a_i^\dagger$ are boson operators obeying $[a_i\nd,a_j^\dagger] = \delta_{ij}$. (Note that these bosons are not simply related to the original hard-core bosons of the quantum problem.)

For large $S$, the square roots can be expanded as power series in $S^{-1}$. While such an expansion should clearly not be expected to give quantitatively accurate results for the case $S=\frac{1}{2}$, we will show that it already captures the expected physics at quadratic order in the operators $a_i\nd$ and $a_i^\dagger$, even at half filling, where the hard-core nature of the bosons is most important.

The classical ground state can be found by applying Eqs.~(\ref{HPstart}) to (\ref{HPend}) to the Hamiltonian $\quH$ and keeping only those terms in the expansion that contain no boson operators. This gives
\begin{multline}
\label{Ham0}
\Ham _0 = N\mu S\cos\theta - S^2\sum_{ij}t_{ij} \sin^2\theta \\
- S^3\sum_{ij\ell} w_{ij\ell} \cos\theta \sin^2\theta\punc{,}
\end{multline}
where $N$ is the number of sites in the lattice. The final term vanishes because $w_{ij\ell} = -w_{ji\ell}$. Let $t_0 = \sum_j t_{ij}$, which can be shown to be independent of $i$ using a combination of the transformations $\xTrans$, $\yTrans$ and $\yReflect$, defined in Section~\ref{SecSymmetries}. The minimum of $\Ham _0$ is given by choosing
\beq{CosTheta}
\cos\theta = -\frac{\mu}{2S t_0}\punc{,}
\eeq
provided that $-2S t_0 < \mu < 2 S t_0$, or equivalently, that the chemical potential lies within the dispersion band. At the transition into the vacuum, the chemical potential goes down through the bottom of the band; within the vacuum phase, $\mu < -2 S t_0$, and the minimum is instead given by $\cos \theta = 1$.

With this choice for $\theta$, the terms in the expansion that are linear in boson operators cancel. The next terms are quadratic in the boson operators and, as usual for the Holstein-Primakoff expansion, include terms of the form $a_i a_j$ and $a_i^\dagger a_j^\dagger$ as well as number-conserving terms such as $a_i^\dagger a_j\nd$. After some algebra, these quadratic terms can be written as
\beq{Ham2}
\Ham _2 = \frac{1}{2}\sum_{ij}
\begin{pmatrix}
a_i^\dagger & a_i\nd
\end{pmatrix}
\begin{pmatrix}
A_{ij}+B_{ij}&C_{ij}-D_{ij}\\
C_{ij}+D_{ij}&A_{ij}-B_{ij}
\end{pmatrix}
\begin{pmatrix}
a_j\nd\\
a_j^\dagger
\end{pmatrix}\punc{,}
\eeq
where
\begin{align}
\label{ABCDdefine1}
A_{ij}& = 2S t_0\delta_{ij} -S t_{ij}(1+\cos^2\theta)\\
B_{ij}& = 2S^2\cos^2\theta\sum_\ell w_{ij\ell}+S^2\sin^2\theta\sum_{\ell}(w_{\ell ij}-w_{\ell ji})\\
C_{ij}& = S t_{ij}\sin^2\theta\\
D_{ij}& = -S^2\sin^2\theta\sum_{\ell}(w_{\ell ij}+w_{\ell ji})\punc{.}
\label{ABCDdefine2}
\end{align}
Note that $A_{ij}$, $C_{ij}$ and $D_{ij}$ are symmetric in their indices, while $B_{ij}$ is antisymmetric. The nonhermitian nature of $\quH$ is therefore reflected in $\Ham _2$: taking the hermitian conjugate of $\Ham _2$ gives $B_{ij}\rightarrow -B_{ij}$ and $D_{ij} \rightarrow -D_{ij}$.

\subsection{Bogoliubov transformation}

The quadratic Hamiltonian $\Ham _2$ can be diagonalized using the standard Bogoliubov transformation of bosons, in which some care must be taken because $\Ham _2$ is nonhermitian. First define the column vector $\boalpha$ so that $\alpha _i = a_i$ for $1 \le i \le N$ and $\alpha _i\nd = a_{i-N}^\dagger$ for $N+1 \le i \le 2N$. The commutation relations become $[\alpha _i\nd,\alpha _j^\dagger] = \eta _{ij}$, where $\boeta$ is a $2N\times 2N$ diagonal matrix with $+1$ for the first $N$ elements on the diagonal, and $-1$ for the remainder. We transform from $\boalpha$ and its conjugate $\boalpha^\dagger$ to $\boalt$ and $\boaltbar$ (which are not hermitian conjugates), by
\beq{Bogoliubov}
\boalpha = \mathbf{V}\boalt \qquad\text{and}\qquad \boalpha^\dagger = \boaltbar \mathbf{W}\punc{.}
\eeq
The matrices $\mathbf{V}$ and $\mathbf{W}$ are not hermitian conjugates, but are instead related to preserve the commutator: $[\tilde\alpha _i,\bar{\tilde\alpha}_j] = \eta _{ij}$, so that $\mathbf{W} = \boeta \mathbf{V}^{-1} \boeta$.

The quadratic Hamiltonian $\Ham_2$ can be written as
\beq{Ham2b}
\Ham _2 = \frac{1}{2}\boalpha^\dagger \boH \boalpha
\eeq
(where a matrix product is implied), so we must choose $\mathbf{V}$ such that $\tilde{\boH} = \mathbf{V}^{-1} \boeta \boH \mathbf{V}$ is a diagonal matrix, giving
\beq{Ham2diag}
\Ham _2 = \frac{1}{2}\boaltbar \boeta \tilde{\boH} \boalt\punc{.}
\eeq
Note that, since neither $\boH$ nor $\boeta \boH$ is hermitian, the elements of $\tilde{\boH}$ are not necessarily real.

Nonetheless, it can be shown\cite{BlaizotRipka} that for every eigenvalue $\epsilon _i$ of $\boeta \boH$, there is a corresponding eigenvalue $-\epsilon _i$. After appropriately ordering the eigenvalues and eigenvectors, one can therefore identify the elements $\tilde\alpha _i$ as annihilation ($1 \le i \le N$) and creation ($N+1 \le i \le 2N$) operators and write
\beq{Ham2final}
\Ham _2 = \sum_{i = 1}^{N} \epsilon _i \bar{\tilde{a}}_i \tilde{a}_i\punc{,}
\eeq
where $\epsilon _i$ are the eigenvalues with positive real parts (and a real constant has been dropped). Note that $\bar{\tilde{a}}_i$ is not the hermitian conjugate of $\tilde{a}_i$, and so that number operator $\bar{\tilde{a}}_i \tilde{a}_i$ is not hermitian. The commutation relations nonetheless ensure that $\tilde{a}_i$ and $\bar{\tilde{a}}_i$ act in a {\it bona fide} Fock space, and that the number operator has real, nonnegative-integer eigenvalues.

In our case, where the Hamiltonian and hence $\boH_{ij}$ are invariant under the translation operators $\xTrans$ and $\yTrans$, the eigenvectors can be assigned to momenta within the reduced Brillouin zone. This contains $\frac{1}{2}N$ points, and we therefore expect two bands, of which only the lower band will be important for the long-range properties with which we are concerned. Let $\at_\kv$ and $\atb_\kv$ be the operators for a state in the lower band with (crystal) momentum $\kv$, and let $\epsilon _\kv$ be the corresponding energy. These lower-band excitations are the phonon modes of the condensate and will therefore have a linear spectrum,
\beq{PhononSpectrum}
\epsilon _\kv = \sqrt{c_x^2 k_x^2 + c_y^2 k_y^2}\punc{,}
\eeq
where $c_x$ and $c_y$ are the phonon velocities.

In terms of the coefficients defined in Eqs.~(\ref{ABCDdefine1}) to (\ref{ABCDdefine2}), $c_x$ and $c_y$ are given by
\begin{align}
c_x^2 &= 2St_0 \sin^2 \theta \sum_i x_i^2 (C_{i0} - A_{i0})\\
c_y^2 &= 2St_0 \sin^2 \theta \Bigg\{ \sum_i y_i^2 (C_{i0} - A_{i0}) \; +\nonumber\\
&\qquad\qquad\qquad\qquad\frac{\left[\sum_j y_j (B_{j0}-D_{j0}) \right]^2}{\sum_\ell (-1)^{y_\ell}(A_{\ell 0} + C_{\ell 0})} \Bigg\}\punc{.}
\end{align}
While the full spectrum is not necessarily real, the assumption that the ground state of the system is a stable superfluid implies that the phonon velocities are real.

\subsection{Correlation functions}
\label{LargeSCorrFunc}

The connected part of the spin--spin correlation function in spin ice, $\C\sub{c}(\rv,\rvp;z)$, for spins at points $\rv$ and $\rvp$ in their respective $(100)$ planes and separated by $z$ in the $[100]$ direction, translates in the hard-core boson language to the time-ordered number--number correlation function. At inverse temperature $\beta$ (which we will later take to infinity), the expectation value of the number operator at site $i$ is
\beq{NumberEV1}
\langle n_{i}\rangle = \frac{1}{\Z}\Tr \left(\T^\beta n_i\right)\punc{,}
\eeq
where $\Z = \Tr \T^\beta$ is the partition function. Note that the operators here refer to the hard-core bosons $b_i$, rather than the Holstein-Primakoff bosons $a_i$. The pairwise correlation function (including disconnected parts) is
\beq{CorrFunc1}
\C(i,j;z) =
\begin{cases}
\frac{1}{\Z}\Tr \left(\T^{\beta-z} n_i \T^z n_j\right) & \text{for $z>0$,}\\
\frac{1}{\Z}\Tr \left(\T^{\beta-|z|} n_j \T^{|z|} n_i\right) & \text{for $z<0$.}
\end{cases}
\eeq
(Note that the relative units for distance in the $[100]$ direction and within the $(100)$ plane are arbitrary, and so we have chosen to take the unit of $z$ as four layers in the $[100]$ direction.)

First observe that exchanging the two sites $i$ and $j$ is, as expected, equivalent to taking $z \rightarrow -z$; we therefore focus on $z>0$.

We rewrite the correlation function using a spectral representation, by diagonalizing the transfer operator as
\beq{DiagonalizeT}
\T = \sum_\stateenergy \ket{\stateenergy} \ee^{-\stateenergy} \bra{\stateenergy}\punc{,}
\eeq
where $\stateenergy$ labels a complete set of eigenstates of the Hamiltonian. This gives
\beq{NumberEV2}
\langle n_{i}\rangle = \frac{1}{\Z}\sum_{\stateenergy} \bra{\stateenergy} n_i\ket{\stateenergy}\ee^{-\beta\stateenergy}
\eeq
and
\beq{CorrFunc2}
\C(i,j;z) = \frac{1}{\Z}\sum_{\stateenergy,\stateenergy'} \bra{\stateenergy} n_i \ket{\stateenergy'} \bra{\stateenergy'} n_j \ket{\stateenergy} \ee^{-\stateenergy(\beta-z)}\ee^{-\stateenergy'z}\punc{,}
\eeq
which reduce, in the zero-temperature limit, to
\beq{NumberEV3}
\langle n_{i}\rangle = \bra{\stateenergy_0} n_i\ket{\stateenergy_0}
\eeq
and
\beq{CorrFunc3}
\C(i,j;z) =
\sum_{\stateenergy} \bra{\stateenergy_0} n_i \ket{\stateenergy} \bra{\stateenergy} n_j \ket{\stateenergy_0} \ee^{-(\stateenergy-\stateenergy_0)z}\punc{,}
\eeq
where $\stateenergy_0$ labels the ground state. The term in the sum with $\stateenergy = \stateenergy_0$ is independent of $z$ and cancels when the connected correlation function $\C\sub{c}$ is calculated:
\begin{align}
\label{CorrFunc3b}
\C\sub{c}(i,j;z) &=\C(i,j;z) - \langle n_i \rangle\langle n_j \rangle\\
&=\sum_{\stateenergy\neq \stateenergy_0} \bra{\stateenergy_0} n_i \ket{\stateenergy} \bra{\stateenergy} n_j \ket{\stateenergy_0} \ee^{-(\stateenergy-\stateenergy_0)z}\punc{.}
\end{align}

The translation symmetry of the Hamiltonian means that we can choose a set of eigenstates $\ket{\stateenergy}$ which are also eigenstates of the translation operators $\xTrans$ and $\yTrans$. In particular, the eigenvalue of the translation operators in the ground state will be $+1$. We let $n_\sublattice{a}$ and $n_\sublattice{b}$ denote the number operators for the sites belonging to each of the two sublattices in the unit cell at $\rv = \zerov$. The translation operators can then be used to give
\beq{CorrFunc4}
\C\sub{c}(i,j;z) =
\!\sum_{\stateenergy\neq \stateenergy _0}\! \bra{\stateenergy_0} n_{\sigma_i} \ket{\stateenergy} \bra{\stateenergy} n_{\sigma_j} \ket{\stateenergy_0} \ee^{-(\stateenergy-\stateenergy_0)z}\ee^{\ii \Kv _{\stateenergy} \cdot \rv_{ij}}
\eeq
where $\sigma _i \in\{\sublattice{a},\sublattice{b}\}$ denotes the sublattice that site $i$ belongs to, $\rv _{ij}$ is the separation of sites $i$ and $j$ within the $(100)$ plane, $z$ is their separation in the $[100]$ direction, and $\Kv _{\stateenergy}$ is the total momentum of the state $\stateenergy$.

The matrix elements $\bra{\stateenergy_0} n_{\sigma} \ket{\stateenergy}$ and $\bra{\stateenergy} n_{\sigma} \ket{\stateenergy_0}$ (which are not complex conjugates, because of the nonhermitian Hamiltonian) are to be calculated consistently within the Holstein-Primakoff expansion. Using the mapping to quantum spins and Eqs.~(\ref{HPstart}) to (\ref{HPend}), we have
\beq{NumberOperatorsInHP}
n_i = - S\cos\theta + \frac{1}{2} + \sqrt{\frac{S}{2}}(a_i\nd+a_i^\dagger) \sin\theta+ \cdots \punc{,}
\eeq
where the omitted terms are at least quadratic in the boson operators. The matrices $\mathbf{V}$ and $\mathbf{W}$ can then be used to express $a_i$ and $a_i^\dagger$ in terms of $\tilde{a}_{\kv}$ and $\bar{\tilde a}_{\kv}$, using \refeq{Bogoliubov}.

The quadratic approximation to the Hamiltonian, $\Ham _2$, has eigenstates with definite numbers of quasiparticle excitations, and the operator $n_i$, which is linear in $\tilde{a}_{\kv}$ and $\bar{\tilde a}_{\kv}$, can only create or annihilate a single quasiparticle. The constant terms in the expansion for $n_i$ will therefore always cancel when calculating off-diagonal matrix elements: these will involve only states $\stateenergy$ with a single quasiparticle. If the momentum of this quasiparticle is $\kv$, then $\Kv _\stateenergy = \kv$, and one can write
\beq{NumberME}
\bra{\stateenergy_0} (a_{\sigma}\nd + a_{\sigma}^\dagger) \ket{\stateenergy} = f_\sigma(\kv)
\;\text{and}\;
\bra{\stateenergy} (a_{\sigma}\nd + a_{\sigma}^\dagger) \ket{\stateenergy_0} = f_\sigma '(\kv)\punc{.}
\eeq

The connected correlation function can then be written in the form of an integral over $\kv$,
\beq{CorrFunc}
\C\sub{c}(i,j;z) = \frac{S}{2}\sin^2\theta 
\int\frac{\dd^2 \kv}{(2\pi)^2} f_{\sigma _i}\nd(\kv) f_{\sigma _j}'(\kv) \ee^{-\epsilon _{\kv} z}\ee^{\ii \kv \cdot \rv_{ij}}\punc{.}
\eeq
Note that, if $f_\sigma$ and $f_\sigma '$ were complex conjugates, as would be the case for a hermitian Hamiltonian, this expression would be symmetric under $i \leftrightarrow j$. The nonhermitian nature of the Hamiltonian is therefore crucial to the directional dependence of the correlation functions, as noted in Section~\ref{EffHam}.

We have so far restricted to $z > 0$; the form for $z < 0$ is given by exchanging $i$ and $j$, or equivalently by exchanging $\sigma _i$ and $\sigma _j$ and taking $\kv \rightarrow -\kv$. The Fourier transform to momentum- and frequency-space is therefore given by
\beq{kCorrFunc}
\C_{\sigma \sigma '}(\kv, \omega) = \frac{S}{2}\sin^2\theta \left[ \frac{f_{\sigma}\nd(\kv) f_{\sigma '}'(\kv)}{\ii \omega + \epsilon_{\kv}} + \frac{f_{\sigma '}\nd(-\kv) f_{\sigma}'(-\kv)}{-\ii \omega + \epsilon_{\kv}} \right]\punc{.}
\eeq

As in the square-lattice case,\cite{Jaubert} one finds $f _\sigma(\kv) \sim \sqrt{|\kv|}$ for small $|\kv|$. The anisotropy in the Hamiltonian leads, in this case, to a dependence on the direction of $\kv$ of the form
\beq{fForm}
f_\sigma(\kv) \sim \frac{\pm\ii v k_y + \epsilon _\kv}{\sqrt{\epsilon _\kv}}
\eeq
(and the same for $f'$), where $\epsilon _\kv$ is the dispersion given in \refeq{PhononSpectrum}, $v$ is a real constant, and the sign of the imaginary part is $+$ when $\sigma$ is in the $\sublattice{a}$-sublattice and $-$ for the $\sublattice{b}$-sublattice.

The characteristic velocity $v$ can be written as
\beq{CharacteristicVelocity}
v = 2St_0 \sin^2 \theta \frac{\sum_j y_j (B_{j0}-D_{j0})}{\sum_\ell (-1)^{y_\ell}(A_{\ell 0} + C_{\ell 0})}\punc{,}
\eeq
in terms of the coefficients defined in Eqs.~(\ref{ABCDdefine1}) to (\ref{ABCDdefine2}).

One therefore finds, for the connected correlation function within the same sublattice,
\beq{CorrFuncSame}
\C_{\sublattice{a}\sublattice{a}}(\kv,\omega) \sim \frac{S}{2}\sin^2\theta  \cdot \frac{c_x^2 k_x^2 + (c_y^2 - v^2)k_y^2 - 2 v k_y \omega}{\omega^2 + \epsilon _\kv ^2}\punc{,}
\eeq
with $v \rightarrow -v$ for $\C_{\sublattice{b}\sublattice{b}}$. For opposite sublattices we have instead
\beq{CorrFuncOpp}
\C_{\sublattice{a}\sublattice{b}}(\kv,\omega) \sim \frac{S}{2}\sin^2\theta  \cdot \frac{c_x^2 k_x^2 + (c_y^2+v^2) k_y^2}{\omega^2 + \epsilon _\kv ^2}\punc{.}
\eeq
It can immediately be seen that the former expression has a spatial asymmetry and hence a preferential direction, while the latter is symmetric in $\pm k_y$.

These expressions are in fact exactly equivalent to those given in Eqs.~(\ref{aaCorrFunc}) and (\ref{abCorrFunc}) in Section~\ref{SecCoulombPhase}, and in Eqs.~(\ref{nnCorrelation1}) to (\ref{nnCorrelation3}) in Section~\ref{SecCondensedPhase}, as can be shown by subtracting $(S/2)\sin^2\theta$ from both. (This term, which is independent of $\kv$ and $\omega$, becomes a delta function in real space and hence has no effect on the long-range behaviour.)

\subsection{Intermediate asymptotics near transition}

As the phase transition is approached from the superfluid, the chemical potential decreases towards the bottom of the band, and \refeq{CosTheta} implies $\cos\theta = 1$ at the transition point. It follows that the phonon velocities $c_x$ and $c_y$ vanish like the square root of the chemical potential as the transition is approached, as noted in Section~\ref{SecCondensedPhase}. There is therefore an intermediate regime of behaviour within the superfluid but close to the transition, where $\sin\theta \ll |\kv| \ll 1$ and the linearized dispersion relation in \refeq{PhononSpectrum} is no longer valid.

From Eqs.~(\ref{ABCDdefine1}) to (\ref{ABCDdefine2}), one finds that $C_{ij}$ and $D_{ij}$ vanish as $\sin \theta \rightarrow 0$, and so the only terms that remain in the quadratic Hamiltonian $\Ham _2$ are those that conserve (Holstein-Primakoff) boson number. The Hamiltonian is therefore diagonalized with a straightforward canonical transformation, giving a dispersion relation of the form
\beq{QuadraticDispersion}
\epsilon_{\kv} = \frac{k_x^2}{2m_x} + \frac{k_y^2}{2m_y}\punc{,}
\eeq
and constant matrix elements $f_\sigma , f_\sigma ' \sim 1 + \Order{|\kv|}$.

The connected correlation function is therefore given by
\beq{kCorrFunnIA}
\C(\kv,\omega) \sim \frac{\epsilon_{\kv}}{\omega^2 + \epsilon_{\kv}^2}\punc{,}
\eeq
with no sublattice dependence, to leading order. After Fourier transforming back to real space, this gives
\beq{rCorrFunnIA}
\C(\rv,z) \sim \frac{1}{|z|} \exp \left(-\frac{m_x x^2 + m_y y^2}{2|z|}\right)\punc{,}
\eeq
the correlation function for a random walk in two dimensions.\cite{Jaubert} This is to be expected near the transition into the vacuum, where the density is small and the correlations at distances shorter than the string separation are controlled by the behaviour of an isolated string. In this random-walk regime, the correlation function no longer reflects the directed hopping of the original Hamiltonian, since a single step is sufficient to randomize the sublattice of the string.

\section{Discussion}
\label{SecDiscussion}

In summary, we have studied spin ice in a $[100]$ magnetic field by mapping the classical statistical system in three spatial dimensions into a quantum model in two spatial dimensions. The thermal phase transition between a Coulomb phase at weak field and an ordered phase at strong field was mapped to a quantum phase transition between a condensed phase of bosons and the vacuum. While one cannot na\"\i vely write a Landau-Ginzburg-Wilson (LGW) theory for the original transition, since neither state is thermally disordered, the quantum transition is described by the standard critical theory for bosons at low density.\cite{FisherHohenberg,Subir}

The quantum Hamiltonian that results from the mapping is strongly interacting, and we have not attempted to find it explicitly. Instead, we showed how symmetry considerations can be used to constrain the general form for the Hamiltonian and for the corresponding continuum action. Using these general forms, we presented two different calculations of the correlations within the superfluid phase, which, in the zero-field limit, agree with previous results for the classical Coulomb phase.\cite{Huse,Isakov1}

While the mapping from quantum mechanical systems in $d-1$ dimensions to classical systems in $d$ dimensions is of course standard, we believe that this is the first example of the use of the reverse mapping to derive a LGW theory for a na\"\i vely non-LGW thermal phase transition. (See Ref.~\onlinecite{Pokrovsky} for an example of an application to an LGW transition in two spatial dimensions.) As noted above, a forthcoming work will present the application of this approach to the ordering transition of close-packed dimers on the cubic lattice,\cite{Alet} which is also believed to exhibit a non-LGW thermal phase transition.

\begin{acknowledgments}
One of us (JTC) thanks P.\ C.\ W.\ Holdsworth, L.\ D.\ C.\ Jaubert, and R.\ Moessner for closely related collaborations. We are grateful to C.\ Castelnovo for helpful comments. The work was supported in part by EPSRC Grant No.\ EP/D050952/1.
\end{acknowledgments}

\appendix*
\section{Symmetries of the continuum theory}
\label{ContinuumTheory}

Here we present a method by which the continuum action presented in Section~\ref{SecContinuumTheory} could in principle be derived directly, clarifying the precise form of time-reversal symmetry in the continuum theory.

The first stage is to express the partition function for the bosonic problem as a coherent-state path integral (with the hard-core constraint enforced by an explicit term added to the Hamiltonian). The quadratic hopping term is then decoupled using a field $\psi _j$ defined on the sites $j$ of the lattice, as in the standard derivation of a continuum theory for the bosonic Hubbard model.\cite{Subir} One can then in principle trace over the bosonic degrees of freedom to give an effective action for $\psi$. Since the field $\psi$ couples linearly to the bosonic field $b$, the effective action is symmetric under phase rotation $\psi\rightarrow\psi\ee^{\ii\theta}$ (for arbitrary real $\theta$) and the condensation transition corresponds to the spontaneous breaking of this symmetry of $\psi$.

In the case of the bosonic Hubbard model, where there is only a quadratic hopping term, decoupling this term leaves a single-site problem, for which tracing over the bosonic degrees of freedom is tractable. This is no longer the case with correlated hopping terms, but the effective action for $\psi$ can still be written in the form
\beq{EffSpsi1}
\ee^{-\Act\sub{eff}[\psi]} = \int \D b \: \ee^{-\Act[b,\psi]}\punc{.}
\eeq
Returning to the Hamiltonian representation, this can be rewritten as
\beq{EffSpsi2}
\ee^{-\Act\sub{eff}[\psi]} = \Tr \left\{\timeordered \exp\left[ -\!\!\int_0^\beta \dd\tau\:\quH\boldsymbol{(}\psi(\tau)\boldsymbol{)}\right]\right\}\punc{,}
\eeq
where $\quH(\psi)$ is the Hamiltonian $\quH$ in which the quadratic, hermitian part of the hopping has been replaced by $\sum_j(\psi _j^* b_j + \psi_j b_j^\dagger)$. The presence of the time-ordering operator $\timeordered$ (for imaginary time $\tau$) makes this form inconvenient to use, and we instead make the time order explicit by writing this as
\beq{EffSpsi}
\ee^{-\Act\sub{eff}[\psi]} = \Tr \lim_{L\rightarrow\infty} \prod _{\iota=1}^L \ee^{-\quH\boldsymbol{(}\psi(\tau _\iota)\boldsymbol{)}\delta \tau}\punc{,}
\eeq
where $\delta\tau = \beta / L$ and $\tau _\iota = \left(\iota - \frac{1}{2}\right)\delta\tau$.

The symmetry properties of the effective action $\Act\sub{eff}$ can be determined from this form of the expression. First, consider a symmetry such as $\yReflect$, defined in Section~\ref{SecSymmetries}, which takes $y\rightarrow -y$ and exchanges the two sublattices, but has no effect on the $z$-direction. This is a symmetry of the Hamiltonian $\quH$, so that we can write $\yReflect\quH\yReflect = \quH$, where $\yReflect$ here denotes the operator representing the transformation (and so $\yReflect^{-1} = \yReflect$). For $\quH(\psi)$, we have $\yReflect\quH(\psi)\yReflect = \quH(\yReflect\psi)$, where $\yReflect\psi$ indicates the result of applying $\yReflect$ to the field $\psi$. Note that here $\psi$ denotes the value of the field at one particular instant, say $\tau _\iota$.

Using \refeq{EffSpsi}, we can therefore write
\beq{EffSpsiRy}
\ee^{-\Act\sub{eff}[\yReflect\psi]} = \Tr \lim_{L\rightarrow\infty} \prod _{\iota=1}^L \ee^{-\yReflect\quH\boldsymbol{(}\psi(\tau _\iota)\boldsymbol{)}\yReflect\delta \tau}\punc{.}
\eeq
The power-series definition of the exponential and the cyclicity of the trace cause the operators to cancel, so that $\Act\sub{eff}[\psi] = \Act\sub{eff}[\yReflect \psi]$.

The same logic applies to particle--hole symmetry at half filling, since $[\quH,\phInvert] = 0$ and so $\phInvert\quH(\psi)\phInvert = \quH(\psi^*)$.

When the Hamiltonian is not particle--hole symmetric, however, the latter identity is no longer true. If the Hamiltonian $\quH$ were hermitian, we would have $\quH(\psi^*) = {[\quH(\psi)]}^\dagger$, but instead taking the hermitian conjugate changes the direction of the directed hopping, and so we have $\yzReflect' \quH(\psi^*) \yzReflect' = [\quH(\yzReflect'\psi)]^\dagger$. We write $\yzReflect'$ rather than $\yzReflect$, since $\psi$ again denotes the value of the field at one particular instant, and so $\yzReflect'$ only inverts the $y$-coordinates.

The analogue of \refeq{EffSpsi} is now
\beq{EffSpsiIy}
\ee^{-\Act\sub{eff}[\yzReflect'\psi]} = \Tr \lim_{L\rightarrow\infty} \prod _{\iota=1}^L \ee^{-\yzReflect'[\quH\boldsymbol{(}\psi^*(\tau _\iota)\boldsymbol{)}]^\dagger\yzReflect'\delta \tau}\punc{,}
\eeq
where the $\yzReflect'$ operators can again be cancelled (and the power-series definition of the exponential yet again used), leaving
\beq{EffSpsiIy2}
\ee^{-\Act\sub{eff}[\yzReflect'\psi]} = \Tr \lim_{L\rightarrow\infty} \prod _{\iota=1}^L {\left[\ee^{-\quH\boldsymbol{(}\psi^*(\tau _\iota)\boldsymbol{)}\delta \tau}\right]}^\dagger\punc{.}
\eeq
The property of the hermitian conjugate that $A^\dagger B^\dagger = (BA)^\dagger$ now allows the product to be reversed; after changing the labelling of the time steps, we have finally $\Act\sub{eff}[\yzReflect'\psi] = \Act\sub{eff}[\tilde\psi^*]$, where $\tilde\psi$ denotes the time-reverse of $\psi$. The operator $\yzReflect$, which effects time reversal as well as $y$-reflection, then obeys the identity $\Act\sub{eff}[\yzReflect\psi] = \Act\sub{eff}[\psi^*]$, as assumed in Section~\ref{SecContinuumTheory}.

\end{document}